# Systematic experimental comparison of particle filtration efficiency test methods for commercial respirators and face masks

Joel C. Corbin*, Greg J. Smallwood, Ian Leroux, Jalal Norooz Oliaee, Fengshan Liu, Timothy A. Sipkens, Richard G. Green, Nathan F. Murnaghan, Triantafillos Koukoulas, Prem Lobo

*Metrology Research Centre, National Research Council Canada, Ottawa K1A 0R6, Ontario, Canada*

## Abstract

Respirators, medical masks, and barrier face coverings all filter airborne particles using similar physical principles. However, they are tested for certification using a variety of standardized test methods, creating challenges for the comparison of differently certified products. We have performed systematic experiments to quantify and understand the differences between standardized test methods for N95 respirators (NIOSH TEB-APR-STP-0059 under US 42 CFR 84), medical face masks (ASTM F2299/F2100), and COVID-19-related barrier face coverings (ASTM F3502-21). Our experiments demonstrate the role of face velocity, particle properties (mean size, size variability, electric charge, density, and shape), measurement techniques, and environmental preconditioning. The measured filtration efficiency was most sensitive to changes in face velocity and particle charge. Relative to the NIOSH method, users of the ASTM F2299/F2100 method have commonly used non-neutralized (highly charged) aerosols as well as smaller face velocities, each of which may result in approximately 10% higher measured filtration efficiencies. In the NIOSH method, environmental conditioning at elevated humidity increased filtration efficiency in some commercial samples while decreasing it in others, indicating that measurement should be performed both with and without conditioning. More generally, our results provide an experimental basis for the comparison of respirators certified under various international methods, including FFP2, KN95, P2, Korea 1st Class, and DS2.

# 1. Introduction

Filtering facepiece respirators, medical masks, and other face coverings are used in a variety of contexts including industrial, healthcare, and public health settings [1] to remove suspended particles from the airstream entering or exiting the wearer's respiratory system. The efficacy of such face coverings depends mostly on the filtration media and the seal with the wearer's face. While leakage must be assessed on an individual-by-individual basis, filtration efficiency can be measured using well-defined, universal test methods. Standardised test methods ensure that repeatable and reproducible results are obtained, and the comparison of these test methods are the focus of this work.

Historically, the National Institute for Occupational Safety and Health (NIOSH) TEB-APR-STP-0059 test method [2] (hereafter referred to as the "NIOSH test method") has been used in both industrial and healthcare [1] contexts to evaluate the sub-micron particle filtration efficiency (PFE) of non-powered N95 filtering-facepiece respirators under US 42 CFR Part 84 (Subpart K), while the ASTM F2299/F2100 method [3] with ASTM F2100 [4] has been used to evaluate the PFE of medical face masks in North America. Other international respirator standards are generally similar to the NIOSH test method with respect to the PFE measurement (Section 4). Other medical face mask standards also exist and are not discussed here (SGS, 2021). For the filtration of larger, supermicron particles such as bacteria, other standards exist that have been addressed previously [5], and are outside of the scope of this work. In addition, reusable barrier fabric face coverings relevant to the COVID-19 pandemic have been tested using a new ASTM PFE method (ASTM F3502-21) as well as analogues [6]. The ASTM F3502-21 method is more closely related to the NIOSH method than the ASTM F2299/F2100 PFE method, as discussed below.

During the COVID-19 pandemic, barrier face coverings such as those addressed by the ASTM F3502-21 method were widely adopted to mitigate the transmission of airborne respiratory diseases [7]. Prior to the COVID-19 pandemic, similar medical masks were also widely used in public in some Asian countries for the same purpose [8]. Although the majority of the mass of respiratory particles is contributed by particles larger than 300 nm aerodynamic diameter [9–11], substantial numbers of smaller particles are also produced [12]. Our understanding of the relative infectiousness of smaller or larger particles is limited relative to our understanding of mask efficacy. Current evidence clearly indicates that smaller particles may be more likely to contain embedded pathogens [13,14]. Conversely, larger particles settle faster [15] and are filtered more easily [11,16]. Since the most penetrating particle size (MPPS) of any filter typically falls in the range 30 nm to 300 nm diameter [17–20], testing face coverings with the similarly sized particles used by the NIOSH and ASTM F2299/F2100 test methods is a logical choice for assessing the PFE of face masks.

The existence of multiple mask-testing standards reflects history rather than a scientific need to differentiate between modes of filtration. Given the current multiplicity of mask testing protocols, a thorough understanding of the significance of any differences between those protocols has become necessary. Such an understanding may guide the eventual international harmonization of mask testing standards.

Many valuable studies have contributed to our understanding of the fundamental mechanisms by which various parameters may influence PFE (e.g. reviews of Refs. [20] and [21]). However, no single study has systematically quantified the influence of key parameters on the PFEs measured under different test methods. This study aims to contribute quantitatively to the understanding of differences between testing procedures. This requires a detailed consideration of the instrumentation, experimental conditions, and aerosol-particle properties involved. We have performed a systematic experimental study of the importance of a range of parameters relevant to the NIOSH 42 CFR Part 84 and ASTM F2299/F2100 test methods, and also discuss the ASTM F3502-21 method afterwards. Our experimental demonstration was performed using a custom-built system which has been validated as capable of producing NIOSH- or ASTM-F2299/F2100-equivalent data [22], as well as with a TSI 8130A automated filter testing instrument. We conclude by discussing the results in the broader context of comparable international standards.

## 2. Methods

### 2.1. Measurement system

The majority of the measurements described herein were carried out using the NRC Particle Filtration Efficiency Measurement System (PFEMS) shown in Figure 1. (See Table 1 for a list of acronyms commonly used in this manuscript.) The PFEMS is described in detail in Smallwood et al. [22] and briefly in the following paragraphs. Some of the measurements in Section 3.5 were performed using a TSI 8130A automated filter testing instrument. The TSI 8130A design is similar to the PFEMS, but the upstream and downstream detectors are light-scattering detectors (photometers) rather than particle counters. The photometers are empirically calibrated to report particle mass. The TSI 8130A measurements were performed following the NIOSH TEB-APR-STP-0059 test method with one exception:

respirators were loaded to approximately 2 mg and not 200 mg for the reasons discussed in Section 3.5. The results from the TSI 8130A were not significantly different from the NRC PFEMS results (Section 2.4).

Table 1. Acronyms used commonly in this manuscript. Acronyms of organizations (ASTM, NIOSH, TSI, NRC) have been excluded.

| **Acronym** | **Definition** |
|---|---|
| CMD | Count median diameter |
| CML | Carboxylate modified latex spheres |
| CPC | Condensation particle counter |
| GMD | Geometric mean diameter |
| GSD | Geometric standard deviation |
| MMD | Mass median diameter |
| MMAD | Mass median aerodynamic diameter |
| MPPS | Most penetrating particle size |
| N95 | NIOSH respirator class |
| PFE | Particle filtration efficiency |
| PFEMS | Particle filtration efficiency measurement system |
| PM | Particulate matter |
| PSL | Polystyrene latex spheres |
| SMPS | Scanning Mobility Particle Sizer |
| VLPM | Volumetric litres per minute |

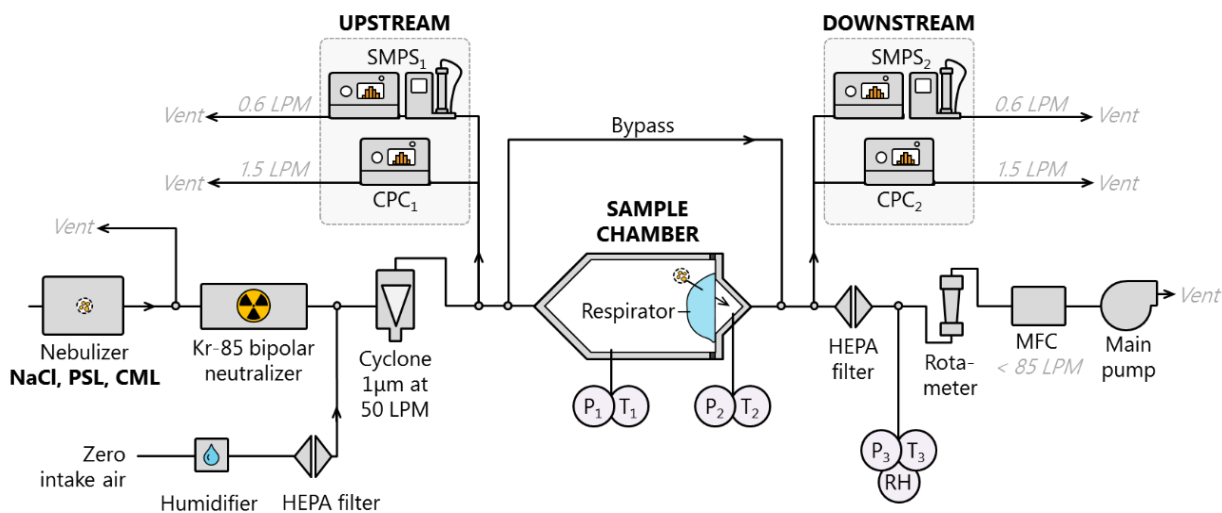

Figure 1. PFEMS configuration setup used for the experiments described in this work.

The PFEMS uses two scanning mobility particle sizers (SMPSs) to measure the number of particles present at a range of sizes (typically 15 nm to 600 nm) upstream and downstream of the sample under test. By selectively analyzing particles of the appropriate mobility diameter (cf. Section 3.1), the PFEMS can provide equivalent results to the ASTM F2299/F2100 method, which uses an optical particle counter but does not precisely distinguish between particle sizes (discussed below). The NIOSH test method, which empirically relates the total light scattering of suspended particles to reference filter samples, can be approximated using the PFEMS via: (i) computing the mass of particles with a given size, (ii) multiplying this mass with the measured number concentration at that size, and (iii) integrating over the size distribution. The uncertainty in such a calculation was minimized using the methods described by Hinds [23] and detailed below. We note that the light-scattering measurements of particle size rely either on knowledge of particle shape and refractive index or an empirical calibration to be accurate: in the ASTM F2299/F2100 method spheres of known refractive index are used, while in the NIOSH method the calibration is empirical.

In the PFEMS, particles were nebulized using a TSI 3076 Collison nebulizer from either a sodium chloride (NaCl) solution or a suspension of polystyrene latex spheres (PSL) in ultrapurified water (resistivity >15 MΩ). In some experiments, carboxylate-modified PSL (CML) spheres were used instead of PSL (Section 3.2). Conductive transport lines (stainless steel or carbon-impregnated silicone tubing) were used throughout the system.

After nebulization, particles were mixed with a high flow of filtered air. The filtered air was humidified as necessary (Table 3) using a Continuous Evaporative Mixer (CEM; Bronkhorst, Netherlands). The combined air flow was controlled by a vacuum pump and mass flow controller at the end of the system to between 12 and 90 volumetric litres per minute (VLPM). A particulate cyclone was placed before the instruments as protection against contamination by coarse particles. The flow in VLPM was converted from the mass flow controller's setpoint in standard litres per minute using temperature and pressure readings from calibrated sensors installed at the points labelled P1, T1, P2, and T2 in Figure 1.

The aerosol flow was passed through filter samples mounted in a custom-made stainless steel sample holder. For N95-type respirators, their edges were glued to the mounting plate, completely sealing the respirator around its perimeter. For medical masks or filter media, the mask was clamped in place. Upstream and downstream of the sample holder, a small portion of the flow was diverted to measure the number concentrations and size distributions of particles using condensation particle counters (CPCs; model 3025A, 3776, and 3788; TSI Inc. USA) and scanning mobility particle sizers (SMPSs). The SMPSs consisted of Differential Mobility Analyzers (DMAs; model 3080 and 3082; TSI Inc.) coupled to similar CPCs. Particles entering the CPCs are grown to easily detectable sizes by condensing water or butanol vapour onto them, before being counted individually by light-scattering. The counts in a specified time interval are normalized by the sample flow rate to obtain particle number concentration $N$. Particle concentration measurements were normalized relative to blank conditions with no sample in the holder and thus are independent of the accuracy of the CPC response. Only the CPC linearity plays a role. The CPCs were used only in their single-particle-counting range and their linearity was verified experimentally [22].

## 2.2. Mass concentrations

A single SMPS measurement provides a histogram of particle number concentrations $dN_i$ for an arbitrary number of bins of mobility diameter $d_m$. When operated properly [24], an SMPS can be used to obtain the total number concentration $N$

$$N = \sum_i dN_i \qquad 1$$

for particles in the SMPS-measured size range. If the number fraction of particles outside of the SMPS size range is negligible, $N$ measured by an SMPS is not different to $N$ measured by a CPC (assuming that $N$ is below the upper limit of the CPC). This was the case in our study.

The SMPS data were used to calculate particle matter (PM) mass concentrations, $M$, in three ways: numerical integration, lognormal fitting, and Hatch-Choate analysis. These three methods are equivalent for ideal data but may differ in practice due to measurement uncertainties such as Poisson counting noise for the largest particle sizes. Numerical integration was performed using [25,26]

$$M = \sum_i dM_i$$

where

$$dM_i = dN_i \cdot m_{p,i} = dN_i \frac{\pi \rho_{\text{eff}}}{6} d_{m,i}^3 \qquad 2$$

where $dM_i$ is the particle mass concentration in the $i$th SMPS bin; $m_p$ is the mass of a particle with mobility diameter $d_m$; $\rho_{\text{eff}}$ is the effective density (discussed in the following paragraphs); and $M$ is the total PM mass concentration.

Lognormal fitting was performed using standard least-square minimization with the geometric mean mobility diameter ($d_{\text{GMD}}$) and geometric standard deviation ($\sigma_g$) from the SMPS measurement used as initial guesses for the position and width of the lognormal. The fitted lognormal distribution was used to calculate integrated particle mass using Equation 2, in the same way as the numerical integration.

Hatch-Choate analysis is based on the assumption of lognormally distributed data. This assumption is valid when the filtration efficiency of a sample changes slowly with size relative to the reference particle distribution, as was the case in all of our samples (see Figure 2 for an example). Hatch-Choate analysis allows the calculation of $M$ from $d_{\text{GMD}}$ and $\sigma_g$ via [23]:

$$M = N \cdot \bar{m} = N \cdot \frac{\pi \rho_{\text{eff}}}{6} \left(d_{\text{GMD}} \exp[1.5 \ln \sigma_g^2]\right)^3 \qquad 3$$

In Equation 3, $\bar{m}$ is the mass of a particle with average mass (which, when multiplied by $N$, gives the total particulate mass). $N$ may be taken from the integrated SMPS data (Equation 1) or from the CPC measurements; these options produced results that were not significantly different.

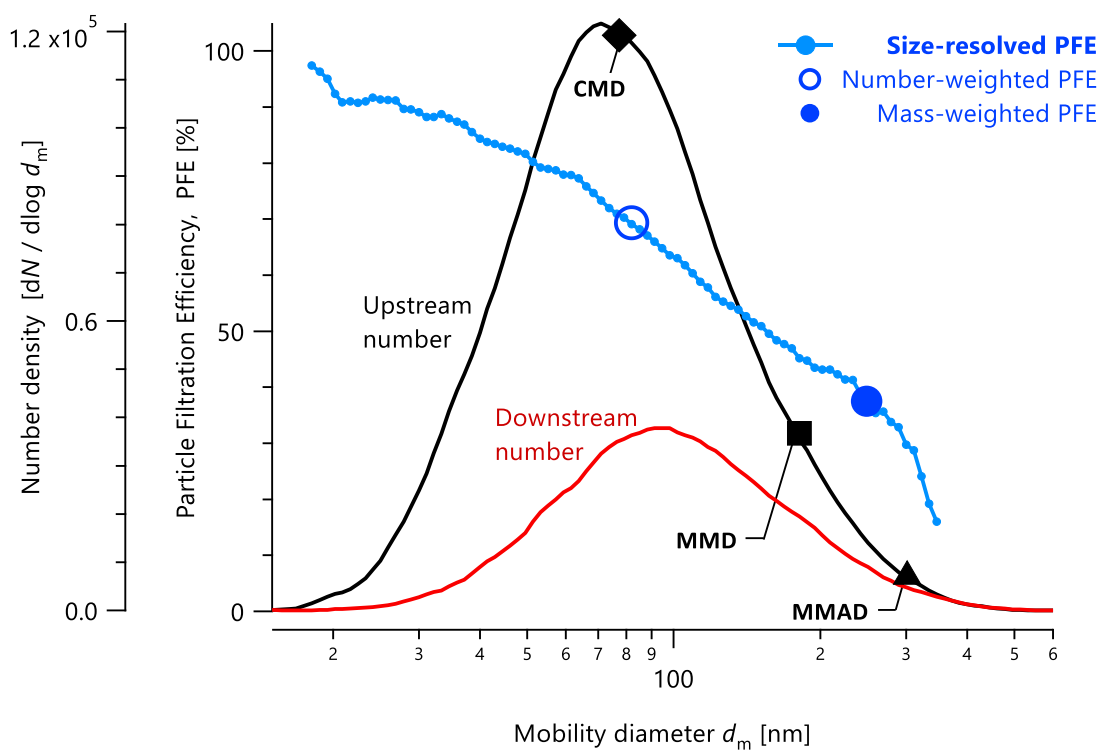

Figure 2. Example size-resolved particle filtration efficiencies (PFEs) measured using the NRC PFEMS system in the NIOSH test method. Summary statistics are labelled on the plot: the overall number- and mass-weighted filtration efficiencies (Equations Equation 1 to 5) and the CMD, MMD, and MMAD (count median diameter; mass median diameter; mass median aerodynamic diameter) for these data. This is an example of a mask with extremely poor performance (30% mass-weighted filtration efficiency), chosen to better illustrate the key features.

The three methods of analysis gave similar results for simulated log-normal, low-noise data. For real experimental data, lognormal fitting was found to be sensitive to the initial guesses of the fitting procedure, while numerical integration became inaccurate when the number concentrations of large (high-mass) particles were low, such as downstream of an efficient filter sample. We therefore used the more robust Hatch-Choate method in all analyses presented in this paper.

We use the mobility diameter $d_m$ in the above equations. It is reasonable to assume that all PSL, CML, and NaCl particles in the measured size range are spherical [27,28]. Under this assumption, the mobility diameter is equivalent to the physical diameter [29], and $\rho_{\text{eff}}$ in Equation 3 is the material density (1050 kg m$^{-3}$ for PSL and CML; 2160 kg m$^{-3}$ for NaCl). If our particles were aspherical, this assumption would have negligible impact on our results since we consider only the ratio of $M$ calculated upstream and downstream of the filter sample.

## 2.3. Filtration efficiency and penetration

Filtration efficiency $E$ is the converse of filter penetration $P$,

$$E = 1 - P \qquad 4$$

Penetration was calculated as

$$P = \frac{c_2}{c_1}\frac{c_1'}{c_2'} = \frac{c_2}{c_1}\cdot\frac{1}{R_{\text{blank}}} \qquad 5$$

where $c_1$ and $c_2$ are the SMPS number (Equation 1) or mass (Equation 2) concentrations upstream and downstream of the sample, respectively, and $c_1'$ and $c_2'$ are analogous, but represent blanks (measurements with an empty sample holder). Therefore, $R_{\text{blank}} = c_2'/c_1'$ accounts for differences in the relative response of the instruments.

The difference between $P$ calculated with respect to number and mass, for Hatch-Choate analysis, can be summarized as (Sipkens et al., in prep.):

$$P_{\text{m}} = \frac{\bar{m}_2}{\bar{m}_1}P_{\text{n}} \qquad 6$$

where $P_{\text{m}}$ and $P_{\text{n}}$ are the penetrations with respect to mass and number and $\bar{m}_1$ and $\bar{m}_2$ are the quantities from Equation 3 for the upstream and downstream measurement locations, respectively.

We measured $R_{\text{blank}}$ at the start and end of each day of measurement and after every 10 samples (when more than 10 samples were measured per day). During routine testing (Section 3.5) this resulted in typically five blank measurements per day, and these measurements showed no temporal trends in their values. This indicates that the instrument drift was negligible, and did not contribute to the uncertainties.

In employing Equation 5, we first converted all concentrations to standard temperature and pressure (293.15 K, 101.325 kPa) using the sensors labelled 1 and 2, respectively. Before and after every day of measurements, we measured zero concentrations by diverting the nebulizer flow and allowing only filtered air through the system, and ensuring that they were negligible relative to the particle counts downstream of high-PFE samples. Therefore, no zero subtraction is included in Equation 5.

Equation 5 can be applied to each SMPS size bin to arrive at a size-resolved penetration in terms of either Equation 2 (particle number per size bin) or Equation 3 (particle mass per size bin), as shown in Figure 2. For size-resolved penetration, there is no meaningful difference between a number or mass basis, since the size bin represents the same group of particles. However, for the number- or mass-weighted average penetration, there is a substantial difference. Since particles of larger diameter have larger mass ($m \propto d^3$), a mass-weighted penetration efficiency is more strongly influenced by the behaviour of larger particles than a number-based one for any given polydisperse test aerosol.

## *2.4. Experimental Uncertainties*

The repeatability of our measured mass-based filtration efficiency (Equation 5) was determined as 0.09 % from 15 repeated measurements of the same respirator.

We also measured 26 respirators with both the PFEMS and a TSI 8130A. Half of the respirators were measured first on the PFEMS, and the other half first on the TSI 8130A. The measured filtration efficiencies ranged from 65 % to 99 %. The difference between the PFEMS and TSI 8130A mass-based filtration efficiencies measurements was not statistically significant and was 1.4 ± 2.2 % (mean and standard deviation). In this work, we propagated and report uncertainties as the standard error (k=1) of repeated measurements.

## *2.5. Single-fibre filtration efficiency model*

Single-fibre filtration efficiency calculations were performed (as shown in Figure 3) to place our results and observations in the context of existing knowledge. We modelled interception, impaction, diffusion (and their total) by reproducing the calculations and assumptions of Ref. [19] (page 116) over a broad size range (10 nm to 1000 nm). We extended these calculations by adding electrostatic capture by dielectrophoretic and coulombic forces following the electrostatic model of Ref. [30]. For this electrostatic model, we additionally assumed an average filter charge density of 1.2 × 10⁻⁴ C m⁻⁴ based on Ref. [31], the physical properties of PSL particles and a polypropylene filter, and singly charged particles for coulombic forces.

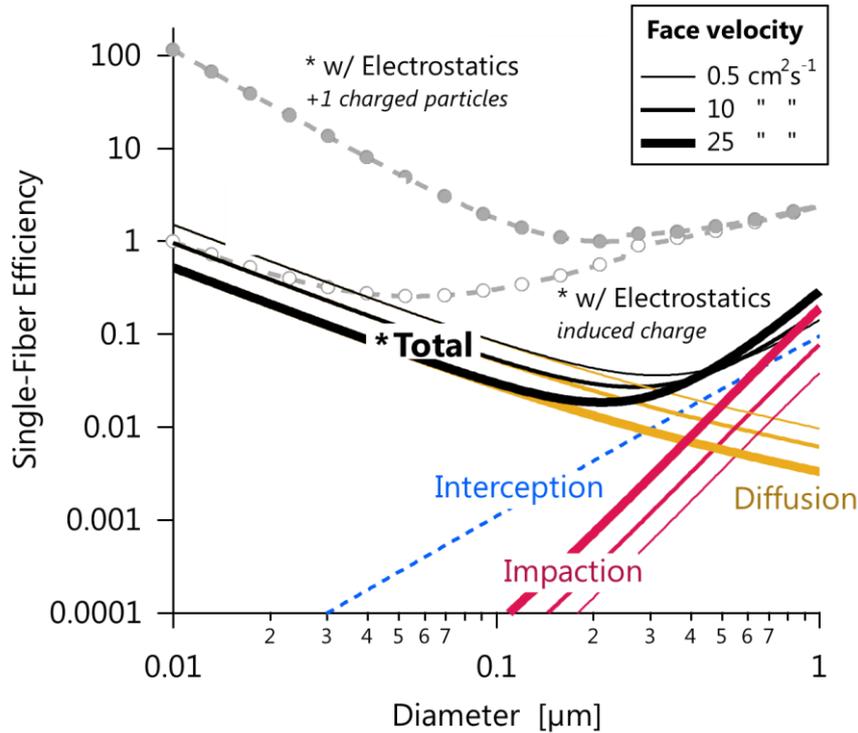

Figure 3. Modelled single-fibre filtration efficiencies for different filtration mechanisms across aerosol-particle sizes relevant to the PFE test methods. The abscissa represents aerodynamic diameter. The physical properties of PSL were assumed for simplicity. The face velocities span the ASTM F2299/F2100 test method at its lowest, NIOSH-like, and maximum face velocities, respectively. Interception capture is independent of face velocity. The "*Total" curve represents an uncharged filter with no electrostatic capture. An electrostatically charged filter may additionally capture uncharged particles by induced-charge (dielectrophoretic) forces (open circles) and charged particles by additional coulombic forces (filled circles), as illustrated for the 10 $cm^2 s^{-1}$ case.

## 3. Results and Discussion

All particle filtration test methods must address certain key experimental variables. At the most basic level, a stream of particles of controlled size must be passed through a filter sample or mask sample at a controlled flow rate, and the filtration efficiency determined from the number or mass of airborne particles upstream and downstream of the filter must be measured. The result of this measurement will be influenced by the electric charge on each particle, the shape and density of particles, and the choice of detection technique for the reasons discussed herein. Table 3 gives an overview of these parameters in the context of the NIOSH and ASTM F2299/F2100 test methods. The following subsections discuss these parameters in detail. Each subsection introduces the fundamental concepts underlying a given parameter, and describes experimental and/or theoretical results to demonstrate the importance of that parameter to test methods. Section 3.2 is an exception to this pattern, since it describes a problem unique to the PSL particles used in the ASTM F2299/F2100 test method.

### 3.1. Size and polydispersity of test particles

The NIOSH and ASTM F2299/F2100 test methods employ particles of different characteristic diameters. These diameters must be compared with care, because there are multiple ways by which to describe the "diameter" of a population of aerosol particles and because these characteristic diameters are often influenced by the polydispersity (range of sizes) of the population [23].

The NIOSH method utilizes NaCl particles lognormally distributed in size with a count median mobility diameter (CMD; the mode of the distribution and equivalent to the geometric mean for lognormal distributions) of 75 nm ± 20 nm and a geometric standard deviation (GSD) not exceeding 1.86. This GSD indicates that 16% of particles are therefore smaller than 40 nm (75 nm ÷ 1.86) and 16% are larger than 140 nm (75 nm × 1.86). The larger particles contribute much more to the total mass than the smaller ones. As a result, although the number distribution has a CMD of 75 nm, the mass median mobility diameter (MMD) of the same particles is larger by a factor of $\exp(3\ln^2 \text{GSD})$ [23]. In this case the equivalent MMD is 240 nm. In the context of particle capture by a filter or within the human respiratory system, the inertia of these particles must be considered by converting this MMD to a mass median *aerodynamic* diameter (MMAD). To a first approximation, MMAD ≈ MMD × $\sqrt{\rho}$, where $\rho$ is the particle's material density in units of g cm$^{-3}$. Specifically, this approximation is only accurate when the aerosol particles are spherical, void-free, and much larger than the mean free path of air [29]. This approximation yields 352 nm MMAD. For an accurate calculation with no approximations [29,32] we calculate an MMAD of 307 nm. Thus, the particles used by the NIOSH standard are "75 nm" from the perspective of diffusion filtration (and the DMA) but "0.3 μm" from the perspective of impaction and interception filtration. Figure 2 illustrates these different diameters with a black diamond (CMD), square (MMD), and triangle (MMAD).

In contrast to the NIOSH method, the ASTM F2299/F2100 method utilizes monodisperse PSL or CML particles. Since these particles are monodisperse (GSD ≈ 1), the CMD and MMD are approximately equal. Since the density of PSL (1.05 g cm$^{-3}$) is close to 1 g cm$^{-3}$ (the reference value used to define aerodynamic diameter; [29]), the MMAD and MMD are also approximately equal. This simplifies the interpretation of particle size in the ASTM F2299/F2100 test. However, in practice, residues and multimers complicate the PSL case, as described in the next section.

The importance of particle size is illustrated by our size-resolved PFE measurements in Figure 2. It is well known that PFE is a function of diameter, with Figure 3 showing the results of single-fibre filtration theory. Smaller particles are captured via diffusion according to their mobility diameters; larger particles are captured via interception and impaction according to their aerodynamic diameters [19]. (Impaction occurs when a flowing particle's inertia brings it into contact with a filter fibre, interception is tangential impaction.) For this reason, size-resolved PFE are often reported in terms of the most-penetrating particle size (MPPS) described in Section 1. Particles smaller or larger than the MPPS are captured more easily by diffusion or impaction and interception [19], respectively. Therefore, the NIOSH and ASTM methods represent particles in the most challenging size range possible. We note that once a particle comes into contact with a filter fibre, van der Waals forces are generally strong enough to prevent its resuspension [33] unless the material is physically agitated [34]. We also note that size-resolved PFEs and the MPPS are much less sensitive to the test aerosol size distribution, which can influence the results of size-integrated PFE measurements like those made in the NIOSH test method and its international analogues [21].

### 3.2. Multimers and residues in PSL aerosols

Figure 4 shows measured size distributions for PSL and CML nebulized from a range of colloid concentrations, as well as a blank sample. As mentioned above, CML is similar to PSL, but rather than being latex particles suspended in a surfactant solution the CML particles are chemically functionalized with polar carboxylate groups at their surface to inhibit their agglomeration in the colloid.

Figure 4a shows, for a variety of concentrations of PSL in aqueous suspension, that the ideal monodisperse size distribution expected for PSL is never achieved. The aerosol always contains substantial numbers of small (about 20 nm in our case) residual particles in addition to PSL. These residual particles reflect the surfactant added to commercial PSL suspensions to inhibit PSL coagulation. Following Ref. [35], we were able to decrease the residual particle number substantially, but not completely, by switching to the surfactant-free CML.

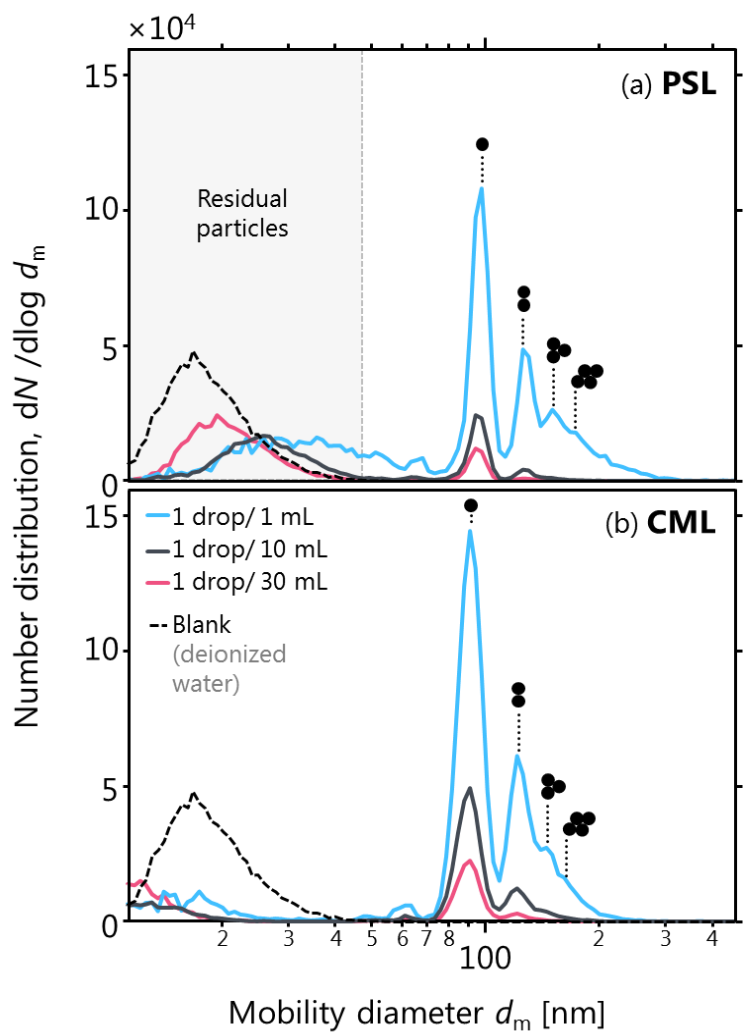

Figure 4. Residues (< 60 nm mobility diameter particles) and multimers (> 100 nm mobility diameter) of PSL suspensions add complexity to aerosols of ostensibly monodisperse particles. Doublets are present even for dilute mixtures. Panel (a) shows PSL with surfactant added to minimize coagulation, Panel (b) shows CML (PSL with surface-bonded carboxylate groups instead of surfactant).

The 20 nm residual particles are estimated to correspond to an impurity concentration of 300 ppm (calculated following Ref. [36]). A tenfold reduction of these impurities (e.g. using reverse osmosis[37]) would decrease the residual particle size by only $10^{-1/3}$ and still produce 12 nm particles. The generation of these residue particles is therefore unavoidable in practice.

The importance of the residue peak can be reduced by increasing the PSL concentration in suspension. However, at high concentrations, multiplets (doublet, triplet, quadruplet, etc.) of PSL become visible in the size distribution. Previous work has confirmed this interpretation using electron microscopy and other measurements [38].

Thus, it is not possible to nebulize a truly monodisperse particle distribution from a PSL suspension. Mobility-size-resolved measurements such as those performed here can be used to selectively analyze the signal from 100 nm particles. For example, when a DMA-CPC combination is used at the standard resolution ($d_m/\Delta d_m$) of 10, the PSL monomer will be adequately resolved from its nearest neighbour (which is a doubly-charged PSL dimer peak). This resolution can be improved up to a resolution of 66 for greater confidence.

Alternatively, a low concentration of PSL can be used in combination with a detector that is insensitive to small residue particles. The latter approach is implicit in the ASTM F2299/F2100 test method, which calls for the use of optical particle counters as detectors. However, the response of such optical particle counters is generally poorer for smaller particles, with performance decreasing rapidly close to 100 nm diameter [39–41], because the light scattering efficiency of small particles decreases rapidly with decreasing particle size. Therefore, while this experimental approach may be insensitive to residue particles, it is also less sensitive to monomer than to dimer PSL/CML particles, resulting in a bias towards dimers in many cases. The magnitude of this bias would depend on instrument design, instrument maintenance, and PSL/CML suspension concentration. Therefore, the sensitivity of optical particle counters used for PFE testing must be verified and checked routinely

One practical solution to the problem of PSL multimers and residues is the use of an SMPS, as in the PFEMS. The data shown in Figure 4 were measured using the PFEMS, and allow the PFE of PSL monomers to be determined with minimal influence of multimers. When using an SMPS for PFE measurements, it is advisable to maintain a relatively low number of multimers to minimize the importance of the SMPS multiple-charging corrections. An alternative solution is the use of an aerodynamic aerosol classifier (AAC)[42] which produces monodisperse aerosols without requiring any corrections.

## 3.3. Face velocity

The face velocity is the speed with which an aerosol passes through a filter and is the ratio of aerosol flow rate to filter surface area. The NIOSH standard is defined in terms of flow rate rather than face velocity because the surface area of respirators may vary between manufacturers. For example, Roberge et al. [43] reported the total inner-layer surface area of 12 N95 respirators as ranging from 108 cm$^2$ to 205 cm$^2$ (mean ± 2 standard deviations, 146 ± 26 cm$^2$). The mean area would result in a mean face velocity of 7.3 ± 1.9 cm s$^{-1}$ for the flow rate of the NIOSH method. However, this mean area is an overestimate; a more accurate calculation would subtract the area of the mask in contact with the wearer's face, which does not contribute to filtration. If this region comprised 10% of the inner-layer area, it would increase the mean face velocity to 8.1 ± 2.0 cm s$^{-1}$. In this section and in Figure 5, we conservatively use a range of 5.4 to 10.1 cm s$^{-1}$, encompassing both of the above estimates, when comparing the face velocities relevant to the NIOSH standard with the ASTM F2299/F2100 standard.

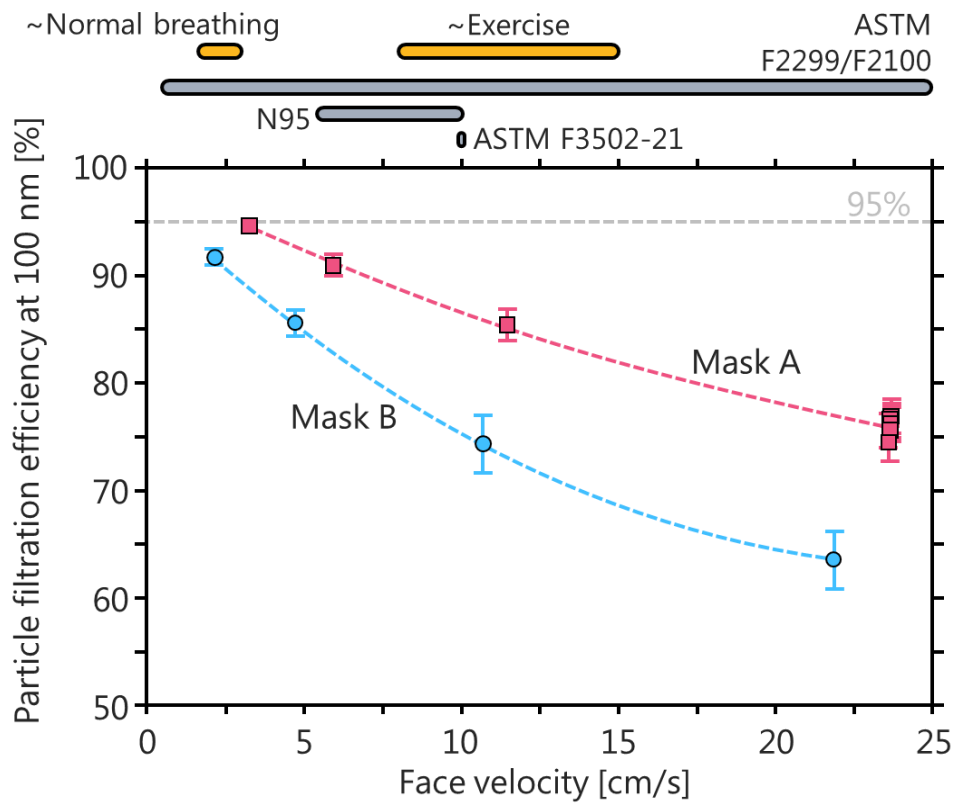

Figure 5. The impact of face velocity on the PFE measured for two medical face masks under the ASTM F2299 test method. The upper grey bars illustrate face velocities (ratio of flow rate to surface area) for the ASTM F2299, ASTM F3502-21, and NIOSH test methods (the range of values indicated for the NIOSH method correspond to the range of N95 surface areas discussed in the text). The upper yellow bars illustrate upper-limit face velocities for normal breathing and exercising flow rates through N95 respirators. The increase in PFE with decreasing face velocity for these 100 nm PSL particles was predicted in Figure 3.

To place the face velocities used in the NIOSH method in a real-world context, we combined the above surface areas with typical breathing flow rates for adults at rest (20 VLPM) and during exercise (100 VLPM) [44]. This gives ranges of 1.6 − 3 cm s$^{-1}$ at rest and 8 − 15 cm s$^{-1}$ during exercise, similar to the face velocities of the NIOSH method. Slightly higher face instantaneous velocities may occur during coughing or sneezing [45–47] at which mask or respirator performance will be reduced, but still substantial [48], and at which point pressure-dependent leak rates may also be relevant. Overall, the face velocities of the NIOSH test method can be considered as reasonable upper limits.

A wide range of face velocities is allowed by the ASTM F2299/F2100 standard, from 0.5 to 25 cm s$^{-1}$. This wide range corresponds to large changes of the measured filter efficiency. Figure 5 shows example data for two medical face masks. When increasing the face velocity from about 2.5 cm s$^{-1}$ to 23 cm s$^{-1}$, the measured efficiency drops from 95% to about 75% for Mask A, and from 92% to 63% for Mask B. This trend is consistent with that predicted from single-fibre filtration theory, shown in Figure 3, and previous measurements [49]. For particles with aerodynamic diameters up to approximately 500 nm, diffusion is the dominant filtration mechanism. At low face velocities, particles are given more time to diffuse to the mask fibres. Considering the ASTM F2100 performance requirement of ≥ 95 % for filtration efficiency for a Level 1 medical mask, this sensitivity to face velocity could mean the difference between accepting or rejecting the masks for medical use. We note that the majority of laboratories use a face velocity of 5 cm s$^{-1}$ when applying the ASTM F2299/F2100 test method at the time of writing (ASTM, private communication, 2020). We also note that this conclusion does not extend to supermicron particles with aerodynamic diameters much larger than 500 nm; these are filtered primarily by impaction (more efficient with increasing face velocity or momentum) and interception (independent of face velocity).

### 3.4. Particle charge and neutralization

Natural aerosol particles are often electrically charged. Nebulized aerosol particles, such as those produced in both NIOSH and ASTM standards, are initially even more highly charged due to the mechanical nebulization process and the fact that the initially nebulized droplets are larger than the dried particles. For example, nebulized NaCl particles of about 700 nm mobility diameter may possess hundreds of charges after nebulization [50]. This value is likely to vary between nebulizers. Therefore, nebulized particles must be "neutralized" to their equilibrium state. The equilibrium state is not zero charge per particle, but that of a Boltzmann distribution [19]. At equilibrium, a population of 100 nm particles has zero, +1, and −1 charges in proportions of approximately 35%, 25%, and 25%; the remaining fraction has multiple charges [51]. We note that not all laboratory-generated particles are highly charged; for example, the vapour-nucleation approach described by Schilling et al.[49] may produce particles closer to a Boltzmann distribution [52].

All neutralizers work by producing a high concentration of positive and negative air ions. Particles passed through the neutralizer interact with these ions until an equilibrium charge distribution is attained. Particle charge plays an important role in filtration efficiency because most high performance respirators utilize electret materials (materials with permanent dipoles) to induce image charges on natural particles and enhance their capture [19]. This mechanism is efficient at all particle sizes. In contrast, mechanical capture (diffusion, impaction, and interception) is not efficient for particles of aerodynamic or mobility diameters between 100 to 300 nm, as shown in Figure 3.

Charge neutralization is required by the NIOSH test method. In contrast, the ASTM F2299/F2100 test method recommends charge neutralization, but does not require it. Historically, the Food and Drug Administration of the United States (US FDA) issued guidance recommending that "unneutralized" particles be used with the previous ASTM test method [53]. Although this previous method has since been withdrawn [5], many laboratories still omit neutralization (ASTM, personal communication, 2020) perhaps because the ASTM test method was not clearly described in a single document [5]. In the past, neutralization was also a practical challenge, as aerosol neutralizers historically contained radioactive substances. This is no longer the case as commercial neutralizers based on electrically produced X-rays are now available.

Figure 6 illustrates the importance of particle charge for 100 nm PSL and NaCl particles. The first bar from the left (grey) in the figure represents an experiment where we removed the neutralizer from our setup. The fraction of uncharged particles was therefore negligible, and all particles were filtered with artificially enhanced efficiency. The second, red bar represents the correctly measured (with neutralization) filtration efficiency. The third and fourth bars from the left represent an experiment where we inserted a DMA after the neutralizer and selected particles with a charge of +1 or -1, respectively. For PSL, the +1 or -1 charge particles were filtered with similar efficiency to the no-neutralizer case. For NaCl, the result was an increase in filtration efficiency relative to the no-neutralizer case, potentially because of the increased importance of easily-filtered multiply-charged (larger) particles due to the higher GSD of the NaCl. These multiply-charged particles represent an instrumental artifact [54] beyond the scope of our demonstrative experiment. Finally, the fifth bar shows an experiment where we inserted an electrostatic precipitator after the neutralizer to remove all charged particles. As expected, the uncharged particles were filtered with the lowest efficiency since electrostatic deposition had played a major role in the initially high filtration efficiency. There was a slight difference in the last experiment between NaCl and PSL, possibly due to the higher density (higher MMAD) of NaCl.

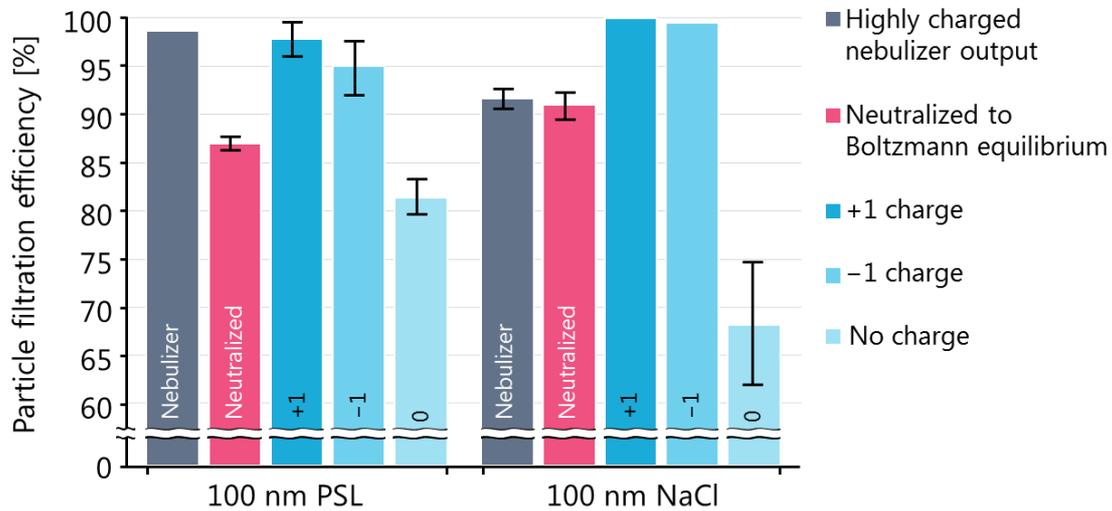

Figure 6. The impact of particle charge on filtration efficiency at 100 nm mobility diameter and 5 cm$^2$ s$^{-1}$ face velocity. Neutral particles (labelled "no charge") are captured least efficiently because they do not induce image charges in the filtration media. The red data labelled "neutralized" represent the efficiency expected for an aerosol at the charge equilibrium (35% zero charge, 25% +1, 25% −1) naturally reached in the atmosphere. The filtration efficiency of neutral NaCl was lower than that of neutral PSL with 100 nm mobility diameter, which may be due to the higher density (higher MMAD) of NaCl.

These experiments demonstrate that any mask testing standard must require that a neutralizer is placed after the particle generation stage. In the context of disease transmission, it might be argued that respiratory droplets might also be highly charged. We are not aware of any studies which have specifically measured the charge of such droplets. However, since respiratory droplets are produced mechanically from saliva and respiratory-tract lining fluid, we expect that the same mechanical processes active in a nebulizer apply to these droplets so that they may be emitted as charged and later neutralized by natural atmospheric ions. Therefore, a laboratory test with neutralized aerosol represents a more conservative result. Moreover, even if natural respiratory droplets were initially charged, any laboratory test must use a neutralizer to obtain well-defined results that are comparable across different test laboratories because the aerosol charge produced by different nebulizers may vary substantially.

*3.5. Preconditioning*

The NIOSH test method calls for the preconditioning of all respirators under humid conditions (85 % ± 5 % RH, 38 °C ± 2.5 °C for 25 h ±1 h) before PFE testing. These environmental conditions are similar to those of exhaled breath and are likely met during prolonged respirator usage. However, these conditions would rarely be met during storage, even in tropical healthcare environments. Consequently, the initial performance of a respirator would be represented by its unconditioned state. After extended usage, the respirator's performance may begin to approach that of its conditioned state. If respirator performance was substantially worse prior to conditioning, then the NIOSH test method would underestimate the true PFE of the device.

Figure 7 shows the impact of preconditioning on 2369 respirators sampled from 221 commercial production lots. The production lots represent a sample of candidate N95 and KN95 respirators tested at the NRC for the Government of Canada in the year 2020. In our samples, the respirator design and manufacturer varied significantly due to the difficulty of procuring a sufficient supply from any given manufacturer at the beginning of the COVID-19 pandemic. Typically 3 to 7 respirators were tested per lot per condition ("unconditioned" or "conditioned"). We used both the PFEMS and a TSI 8130A system for these measurements and observed no significant difference between the two systems for the mass-weighted PFE (Section 2) reported in this analysis. We averaged the data from each lot and calculated uncertainties as the standard error of the mean.

The ordinate of Figure 7 shows the change in PFE upon conditioning. The abscissa shows the initial PFE, representing respirator performance when first donned. The left-centre shaded area (green parallelogram) indicates that the NIOSH test method results for 163 of our 221 respirator lots were not significantly affected by conditioning. (Other measurements, such as gravimetric weighing [19], may still be affected by conditioning.) However, samples from 3 lots would have met the NIOSH requirement (PFE > 95%) after conditioning (top-centre shaded trapezoid), yet failed to meet the standard without conditioning. The opposite was also observed, where samples from 3 lots would have clearly failed with conditioning, but passed without conditioning. Several other cases are apparent in the figure at the pass/fail boundary for the two scenarios; we have not focused on these boundary cases because the definition of a statistically significant failure is beyond the scope of this work.

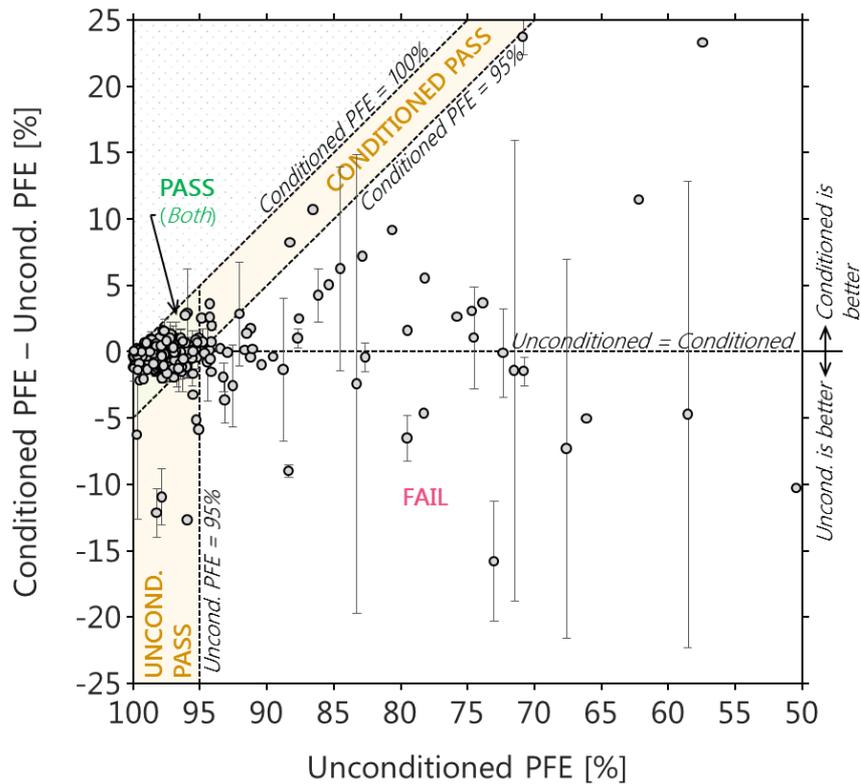

Figure 7. The impact of NIOSH-method preconditioning on respirator particle filtration efficiency (PFE). The ordinate shows the increase (positive) or decrease (negative) in PFE due to conditioning. Overall, 163 of 221 production lots would pass regardless of conditioning (green shading), while 3 clearly pass only if unconditioned (yellow shading, lower) and 3 others clearly pass only if conditioned (yellow shading, upper curve). Several other cases are at the pass/fail boundary for the conditioned-only scenario and their statistical treatment is beyond the scope of this work.

For the lot in Figure 7 with the largest positive effect of conditioning (highest ordinate value), we performed an additional experiment to determine the timescale of this effect. Figure 8 shows that this impact from conditioning required more than 6 h to be observed, but was reversible. (For practical reasons, we were unable to test conditioning periods longer than 6 h and less than 24 h.) Therefore, using this respirator for an 8 hour work shift would result in suboptimal protection for at least half of the shift. We emphasize that this example was an outlier in Figure 7, and is not typical of the evaluated respirators. However, any test method should identify such outliers in order to be effective, as a single outlier may represent many millions of units released to the market. Therefore, the most conservative PFE test method should require testing both with and without conditioning, to represent both the initial and long-term performance of the respirator. The conditioning of respirators is already specified in the FFP2 EN 149-2001, GB2626-2019, and KMOEL-2017-64 test methods (see Table 3 and Section 4.1).

The pressure drop across the majority of respirators was unaffected by conditioning (data not shown). For only one outlier, this pressure drop increased upon conditioning.

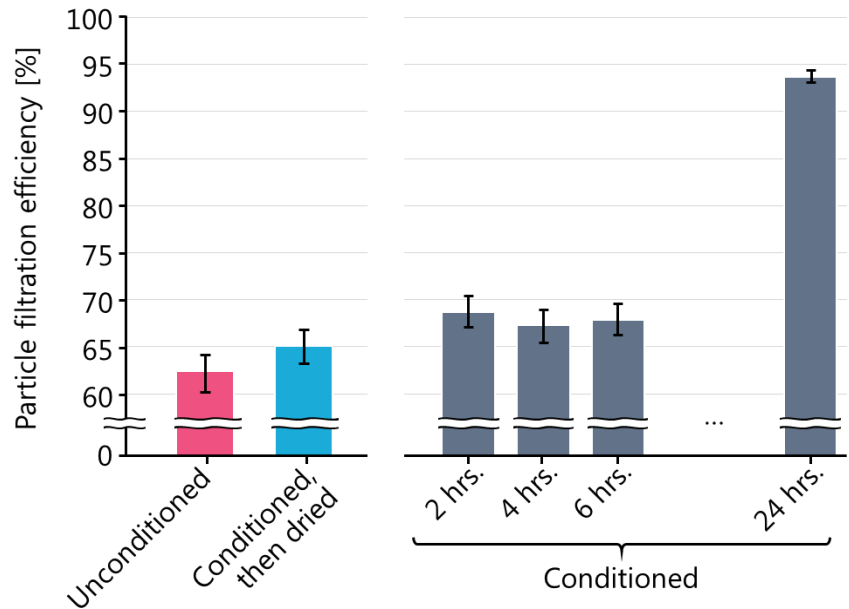

Figure 8. PFE response of one lot of respirators to conditioning over time and to subsequent drying. This lot had the largest difference between unconditioned and conditioned filtration efficiency in our 201-lot data set.

Table 2. Comparison of the NIOSH test method with similar international test methods. Flow resistance parameters have been simplified to maximum flow resistance, see text. PM: particulate matter.

| Certification | N95 | FFP2 | KN95 | P2 | Korea 1st Class | DS2 |
|---|---|---|---|---|---|---|
| Performance standard | US 42 CFR, Part 84 | FFP2 EN 149-2001 | GB2626-2019 | AS/NZ 1716:2012 | KMOEL-2017-64 | JMHLW/214, 2018 |
| Country of origin | USA | EU | China | Australia/NZ | Korea | Japan |
| Particle material[a] | NaCl | NaCl, oil | NaCl | NaCl | NaCl, oil | NaCl |
| Particle size, CMD[b] | 75 nm | 20 nm to 2 μm | 75 ± 20 nm | 20 nm to 2 μm | 40 nm to 1 μm | 60 nm to 1 μm |
| Particle size, MMAD[c] | 300 nm | 600 nm | 300 nm | 300 to 600 nm | 600 nm | n.s. |
| Particle polydispersity | GSD < 1.86 | n.s.[d] | GSD < 1.86 | n.s. | n.s. | GSD < 1.8 |
| Flow rate [VLPM] | 85 | 95 | 85 | 95 ± 2 | 95 | 85 |
| Flow resistance[e] [Pa] | < 245 | < 240 | < 250 | < 120 | < 240 | < 50 |
| Equilibrium charging | Yes | No | Yes | No | No | No |
| Measurand | PM[c] mass | PM mass | PM mass | PM mass | PM mass | PM mass |
| Measurement RH[f] | 30 ± 10 % | n.s. | 30 ± 10 % | n.s. | n.s. | n.s. |
| Mass loading | 200 mg | 120 mg | 200 ± 5 mg | n.s. | n.s. | 100 mg |
| Preconditioning? | Yes (*Humid*[g]) | 3 Yes (*Hot*[h], *Cold*[i]), 3 No | 10 Yes (*Humid, Hot, Cold*), 5 No | No | 5 Yes (*Humid*), 5 No | No |

[a]In all methods the oil used is paraffin oil. [b]Count median mobility diameter. [c]Mass median aerodynamic diameter. [d]n.s.: not specified. [e]Maximum resistance to inhalation or exhalation, whichever is smaller; corresponding flow rates vary between standards[59]. [f]Relative humidity. [g]Humid defined as: condition at 85 % ± 5 % RH, 38 °C ± 2.5 °C for 25 h ±1 h (US 42 CFR, Part 84) or 24 ± 1 h (all others). [h]Hot defined as: dry atmosphere at 70 °C ± 3 °C for 24 h ±1 h, then room temperature for ≥ 4 h. [i]Cold defined as: –30 °C ± 3 °C for 24 h ±1 h, then room temperature for ≥ 4 h.

### 3.6. Mass loading

The mass loading onto a mask or respirator is the amount of PM deposited onto the filter, and is calculated as the product of PM concentration, flow rate through the respirator, and time duration. The ASTM F2299/F2100 test method does not address mass loading, while the NIOSH test method calls for the loading of 200 mg of NaCl onto the respirator. For those respirators which we did load with 200 mg of NaCl, we often observed pressure drops increase beyond the maximum allowance resistance of the NIOSH method. Such a high pressure drop would reduce the breathability of the respirator, causing substantial discomfort for the wearer. This effect is independent of the particle filtration efficiency.

The extreme PM loadings represented by a 200 mg loading are not representative of the healthcare environment nor even most workplaces. For example, 8 h of respirator usage in an extremely polluted urban environment results in a mass loading of < 5 mg for the upper-limit at-rest breathing rate (20 VLPM) given above. This value was calculated assuming 500 µg PM m$^{-3}$, representing an extreme air pollution event ( <1 km visibility) in Shanghai, China [55] or an extremely dirty subway system [56,57]. Conversely, a respirator would need to be reused on 42 shifts of 8 h duration before reaching the 200 mg limit. To reach 200 mg loading in one 8 h shift, 4160 µg PM m$^{-3}$ is required (conditions which can be visualized as comparable to an extreme dust storm; [58]). In a hospital, where 20 µg PM m$^{-3}$ can be assumed, over 1000 shifts would be required. In short, 200 mg of loading would be achieved only in environments where PM loadings are high enough to be visible to the naked eye, as may occur close to significant aerosol sources.

Based on the above considerations, we adjusted the NIOSH method to target only 2 mg of loading (achieved after 1 minute of testing; after which samples were measured for an additional 4 minutes to provide additional statistical information) instead of 200 mg (~2 hour test duration) during our TSI 8130A measurements. This adjustment was also motivated by our need to provide rapid information for the healthcare response to COVID-19. We recommend that the 200 mg loading requirement be reconsidered for healthcare and any other environment where PM loadings are not extremely high.

### 3.7. Measurement techniques

The NIOSH test method calls for measurements of the integrated light scattered by an ensemble of particles to be used to quantify PM mass. The ASTM F2299/F2100 test method calls for optical single-particle counters, which detect individual pulses of scattered light. Fundamentally, these techniques are both capable of providing the required information and should be evaluated in the context of specific interferences or inaccuracies, as discussed in Section 3.2 regarding optical single-particle counters and Ref. [21] regarding light-scattering techniques. Moreover, any revision to these test methods should allow for the use of equivalent techniques such as the SMPS method used here.

## 4. Other standards

### 4.1. Respirator test methods

Our study focused on the NIOSH test method, US 42 CFR Part 84, used for N95 respirator certification. There are other international standards for respirator testing and certification, such as FFP2, KN95, P2, Korea 1st Class, or DS2 respirators, with the characteristics of these standards summarized in Table 3. To a first approximation, these standards are all similar. A respirator that outperforms the requirements of one standard is likely to meet the minimum requirements of the others. However, based on the discussion above, it is apparent that cases where requirements are only barely met may pass the test method of one standard but not another.

For example, while the N95 certification requires preconditioning only, the P2 and DS2 certifications require no preconditioning. The FFP2, KN95, and Korea 1st Class certifications require both preconditioned and unconditioned tests, and vary in their definitions of conditioning. In rare cases (6 of 221 lots, in our data), some respirators clearly performed better with or without conditioning (Section 3.5). Those respirators which performed better with conditioning might obtain N95 certification, but not FFP2, KN95, or Korea 1st Class certification. Our data set indicates that the likelihood of this occurrence is very low for respirators with ≤ 1 % penetration (≥ 99 % filtration efficiency).

An equilibrium charge distribution (neutralization) is not required for FFP2, P2, Korea 1st Class, or DS certification testing. Consequently, a respirator tested under these methods may report a higher filtration efficiency than under the N95 or KN95 protocols (Section 3.4).

Table 3. Summary of the parameters which differ between the NIOSH and ASTM F2299/F2100 test methods. CMD: count median diameter. MMAD: mass median aerodynamic diameter. PSL: polystyrene latex spheres. PFE: particle filtration efficiency. GSD: geometric standard deviation.

| Property | NIOSH 42 CFR 84 | ASTM PFE F2299/F2100 | Section of this manuscript |
|---|---|---|---|
| Particle material | NaCl | PSL | 3.1 |
| Particle size, CMD | 75 nm | 100 nm | 3.1 |
| Particle size, MMAD | 300 nm | 100 nm | 3.1 |
| Particle polydispersity | GSD < 1.86 | Monodisperse | 3.1 |
| Face velocity[c] | ≈5–10 cm s$^{-1}$ | 0.5 – 25 cm s$^{-1}$ | 3.2 |
| Equilibrium charging[d] | Yes | Recommended | 3.4 |
| Flow resistance (Pressure drop) | ≤ 245 Pa | < 353 Pa | 3.5 |
| Preconditioning? | Yes *(Humid)*[e] | No | 3.5 |
| Mass loading | 200 mg | not specified | 3.5 |
| Measurement technique | Total light scattering | Single particle counting | 3.7 |
| Measurand | Particulate mass per m$^{-3}$ air | Particle count per m$^{-3}$ | 3.7 |
| Relative humidity | 30 ± 10 % | 30 – 50 % | – |
| Designed for | Respirators | Medical masks | – |
| Target efficiency | ≥ 95%[f] | ≥ 95%[g] | |

[c]*For N95 respirators the flow rate, not face velocity, is specified. Our quoted range is the 95% confidence interval from N95 surface areas measured by Roberge et al.* [43] *as described in the text.* [d]*Neutralizing to a Boltzmann charge equilibrium state is particularly necessary after producing the test aerosol with a nebulizer.* [e]*Preconditioning at 85 % ± 5 % RH, 38 °C ± 2.5 °C for 25 h ± 1 h.* [f]*For N95 respirators.* [g]*For Level 1 barrier. Level 2 and 3 require ≥ 98%.*

The particle size specified for the different test methods vary somewhat. Also, some test methods also require the use of paraffin oil particles as well as NaCl particles, (which is less relevant in a healthcare context). This difference would become more important at extremely high mass loadings, and be of secondary importance at low loadings. The flow rate also varies between test methods, but by less than 12%. Since the surface area of commercial respirators varies by more than 12% (Section 3.3), this is unlikely to have a major impact (Figure 5).

One dimension where the test methods differ substantially is in their specification of the inhalation and exhalation pressure resistances [59]. In particular, the FFP2 standard defines this measurement at a higher flow rate than the filtration efficiency measurement, making it more challenging.

### 4.2. Barrier face coverings during COVID-19

Early in the COVID-19 pandemic, the importance of universal mask wearing was recognized [e.g. 8], leading to the development of the ASTM F3502-21 standard for barrier face coverings. In contrast to ASTM F2299/F2100, the ASTM F3502-21 standard focused on filtration performance, comfort, and reusability.

ASTM F3502-21 was developed as a modification to the NIOSH test method (42 CFR Part 84, subpart K) rather than a modification to the ASTM F2299/F2100 test method. It differs from the NIOSH test method in the following ways. The method specifies 10 samples are tested unused and 10 additional samples are tested after the maximum number of laundering cycles for reusable masks. All samples are to be preconditioned in the same manner as the NIOSH test method. A narrow range of face velocities, $10 \pm 0.5$ cm s$^{-1}$ but with flow rates not exceeding 85 VLPM, is specified to avoid the issues discussed above (Section 3.3). Two target filtration efficiencies of ≥ 20 % (Level 1) or ≥ 50 % (Level 2) are specified, with corresponding flow resistances of ≤ 147 Pa and ≤ 49 Pa. These lower flow resistances correspond to greater comfort for the wearer and reduced likelihood of leakage.

## 5. Summary

The NIOSH CFR 42 Part 84 and ASTM F2299/F2100 PFE test methods differ in various aspects of their experimental design. These differences fall into three categories: those which affect the physical phenomena underlying filtration, those which reflect the measurement itself, and those which reflect sample conditioning. In this manuscript, we performed systematic experiments to test the impact of each of these differences on the results obtained by either method and discussed the aerosol physics behind the reasons for any differences.

In terms of the physical phenomena, the major differences between the two test methods are that the ASTM F2299/F2100 test method does not explicitly require charge neutralization (but should), allows for a wider range of face velocities (which should be constrained), and uses smaller particles (which present a slightly more challenging test compared to the NIOSH test method). Since filtration efficiency is more sensitive to face velocity than to particle size for this range of values, the ASTM F2299/F2100 test method is likely to report lower penetrations than the NIOSH test method when performed at low face velocities. These comments do not apply to the new ASTM F3502-21 barrier face covering test method, which is more similar to the NIOSH test method than the ASTM F2299/F2100 test method.

In terms of the measurement technique, the NIOSH test method recommends a light-scattering detector calibrated to mass whereas the ASTM F2299/F2100 test method recommends an optical particle counter (OPC). (The new ASTM F3502-21 test method allows for equivalent measurements.) Although OPCs measure number rather than mass, this difference should not be over-interpreted because there is no difference between the number- or mass-based filtration efficiency when truly monodisperse PSL is used. Much more importantly, modern OPCs detect particles smaller than roughly 200 nm with reduced efficiency and are likely to be cross-sensitive or biased towards the multimers (aggregates of two or more particles) that are common in nebulized PSL aerosols. On the other hand, condensation particle counters (CPCs) which overcome this efficiency issue by magnifying particle size prior to detection, are cross-sensitive to the ubiquitous solution residues observed even when nebulizing ultra-pure water. The limitation of OPCs may be addressed with careful calibration, or by replacing the OPC with an optical particle sizer capable of differentiating between monomers and multimers. The limitations of CPCs may be addressed by incorporating a DMA or other particle classifier upstream of the CPC, as in the SMPS configuration used in this work or in an AAC configuration. Future test methods should explicitly allow for alternative measurement methods such as the SMPS method used here.

In terms of conditioning, the ASTM F2299/F2100 test method does not include preconditioning (the ASTM F3502-21 test method does). The NIOSH method requires a long 25 h ± 1 h preconditioning in humid air which is not representative of respirator storage conditions before an 8 h shift, and is much longer than the duration of a typical shift. Conditioning requirements differ for other international NIOSH-like respirator test methods, with some including measurements of both unconditioned and conditioned respirators to ensure representativeness of real-world use conditions. We recommend that future new or modified test methods adopt this practice.

In terms of mass loading, the NIOSH method requires a high 200 mg particle loading which is representative of extreme industrial conditions with PM loadings high enough to limit visibility. This loading is not appropriate for many of the contexts where respirators have been adopted, such as in healthcare.

Overall, there is no physical reason why distinct test methods must be used for the NIOSH and ASTM contexts, and their international equivalents shown in Table 3. The new ASTM F3502-21 test method has already moved towards a harmonized test method by being designed to resemble the NIOSH test method more closely than the ASTM F2299/F2100 test method. Most international respirator standards are also similar to the NIOSH test method and to one another, to a first approximation. For contexts where less stringent mask or respirator performance is required, future applications may consider lowering the minimum required filtration efficiency rather than adjusting the test method itself.

## Acknowledgements

We thank the entire NRC COVID-19 Respirator Testing Team for their contribution to the measurements discussed here: Adam Willes, Adrian Simon, Aiden Korycki-Striegler, Ali Ghaemi, Allison Sibley, Amor Duric, Amr Said, Anabelle Bourgeois, Andason Cen, Andre Cantin, Andrew Oldershaw, Apoorv Shah, Brett Smith, Bryan Muir, Cam Lebrun, Cindy Jiang, Chantal Prévost, Christina Brophy, Dan Clavel, Dave Angelo, David Kennedy, Deval Patel, Doug Mackenzie, Douglas McIntyre, Erhan Dikel, Flamur Canaj, Gang Nong, Garnet McRae, Getho Eliodor, Greg Nilsson, Harold Parks, Isabelle Rajotte, James Renaud, James Saragosa, Jason Brown, Jean Dessureault, Jeff Tomkins, Jeremy Macra, Joshua Marleau-Gillette, Kari McGuire, Kimberly Moore, Krystal Davis, Laura Forero, Lyne St-Cyre, Marilyn Azichoba, Mario Carrière, Mark Vuotari, Michel Levesque, Michael Ryan, Mladen Jankovic, Nicholas Wise, Nikolas Angelo, Ovi Mihai, Peter Hanes, Richard Agbeve, Samuel Camiré, Simon-Alexandre Lussier, Sonia Mutombo, Stacey Lee, Stephane Lapointe, Stephanie Gagné, Steve Kruithof, Thierry Lavoie, Tomi Schebywolok, Tyler Hunter, Vicki Wang, Xigeng Zhao, Yanen Guo.

We thank our colleagues at the Public Health Agency of Canada, Joe Tanelli, Riccardo Santopietro, Denis Laframboise, and Erle Higgins. We also thank colleagues at Health Canada, Carleton University, and the University of Alberta who loaned some instruments and components of the PFEMS at the early stages of this work.

## Author Contributions

J.C.C. and G.J.S. designed research; G.J.S., I.D.L., J.N.O., F.L., T.K. performed experiments; P.L., R.G.G., G.J.S., I.D.L. administered project; J.C.C., G.J.S., N.F.M., R.G.G., T.K., T.A.S. analyzed and interpreted data; and J.C.C. wrote the manuscript with input from all co-authors.

## Funding

This work was funded by the Public Health Agency of Canada and by Pillar 4 of the National Research Council (NRC) Pandemic Response Challenge Program.

## References

1. Radonovich, L. J., Perl, T. M., Davey, V. & Cohen, H. Preventing the Soldiers of Health Care From Becoming Victims on the Pandemic Battlefield: Respirators or Surgical Masks as the Armor of Choice. *Disaster Med. Public Health Prep.* **3**, S203–S210 (2009).

2. NIOSH. Procedure No. TEB-APR-STP-0059, Revision 2.0. Determination of particulate filter efficiency level for N95 series fiters against solid particulates for non-powered, air purifying respirators standard testing procedure (STP). (2007).

3. ASTM. ASTM F2299-03. Standard test method for determining the initial efficiency of materials used in medical face masks to penetration by particulates using latex spheres. (2003). doi:10.1520/F2299_F2299M-03R10

4. ASTM. ASTM F2100-20, Standard Specification for Performance of Materials Used in Medical Face Masks. (2020). doi:10.1520/F2100-20

5. Rengasamy, S., Shaffer, R., Williams, B. & Smit, S. A comparison of facemask and respirator filtration test methods. *J. Occup. Environ. Hyg.* **14**, 92–103 (2017).

6. SGS. Protective Face Mask Testing & Certification. Available at: https://www.sgs.com/en/campaigns/protective-face-masks. (Accessed: 3rd May 2021)

7. Howard, J. *et al.* An evidence review of face masks against COVID-19. *Proc. Natl. Acad. Sci.* **118**, e2014564118 (2021).

8. Feng, S. *et al.* Rational use of face masks in the COVID-19 pandemic. *Lancet. Respir. Med.* **2**, 2019–2020 (2020).

9. Morawska, L. *et al.* Size distribution and sites of origin of droplets expelled from the human respiratory tract during expiratory activities. *J. Aerosol Sci.* **40**, 256–269 (2009).

10. Johnson, G. R. *et al.* Modality of human expired aerosol size distributions. *J. Aerosol Sci.* **42**, 839–851 (2011).

11. Rogak, S. N. *et al.* Properties of materials considered for improvised masks. *Aerosol Sci. Technol.* (2020). doi:10.1080/02786826.2020.1855321